\documentclass[fleqn,usenatbib]{mnras}

\usepackage{newtxtext,newtxmath}

\usepackage[T1]{fontenc}

\DeclareRobustCommand{\VAN}[3]{#2}
\let\VANthebibliography\thebibliography
\def\thebibliography{\DeclareRobustCommand{\VAN}[3]{##3}\VANthebibliography}

\usepackage{graphicx}	
\usepackage{amsmath}	
\usepackage{caption}    
\usepackage{multirow}
\usepackage{orcidlink}

\def\ct{[CII]}
\def\lya{Ly$\alpha$}

\title[Studying {\ct} Emission in Low-mass Galaxies at $z\sim7$]{Studying {\ct} Emission in Low-mass Galaxies at $z\sim7$}

\author[Glazer et al.]{Kelsey Glazer$^{1}$\orcidlink{0000-0002-4453-5870}, 
Maru$\Breve{\textrm{s}}$a Brada$\Breve{\textrm{c}}$ $^{1,2}$\orcidlink{0000-0001-5984-0395},
 Ryan L. Sanders $^{1,8}$\orcidlink{0000-0003-4792-9119},
Seiji Fujimoto $^{3}$\orcidlink{0000-0001-7201-5066},
Patricia Bolan $^{1}$\orcidlink{0000-0002-7365-4131},
\newauthor
Andrea Ferrara$^{4}$\orcidlink{0000-0002-9400-7312},
Victoria Strait $^{5,6}$\orcidlink{0000-0002-6338-7295}, 
Tucker Jones $^{1}$\orcidlink{0000-0001-5860-3419},
Brian C. Lemaux $^{1,7}$\orcidlink{0000-0002-1428-7036},
Livia Vallini$^{9}$\orcidlink{0000-0002-3258-3672},
\newauthor
Russell Ryan$^{10}$\orcidlink{0000-0003-0894-1588}\\
$^{1}$Department of Physics and Astronomy, University of California, Davis, 1 Shields Ave, Davis, CA 95616, USA\\
$^{2}$University of Ljubljana, Department of Mathematics and Physics, Jadranska ulica 19, SI-1000 Ljubljana, Slovenia\\
$^{3}$Department of Astronomy, The University of Texas at Austin, Austin, TX 78712, USA\\
$^{4}$Scuola Normale Superiore,  Piazza dei Cavalieri 7, 50126 Pisa, Italy\\
$^{5}$Cosmic Dawn Center (DAWN), Denmark\\
$^{6}$Niels Bohr Institute, University of Copenhagen, Jagtvej 128, DK-2200 Copenhagen N, Denmark\\
$^{7}$Gemini Observatory, NSF's NOIRLab, 670 N. A'ohoku Place, Hilo, Hawai'i, 96720, USA\\
$^{8}$Department of Physics and Astronomy, University of Kentucky, Lexington, KY 40506, USA\\
$^{9}$INAF-Osservatorio di Astrofisica e Scienza dello Spazio, via Gobetti 93/3, I-40129, Bologna, Italy\\
$^{10}$Space Telescope Science Institute, 3700 San Martin Drive, Baltimore, MD 21218, USA
}
\date{Accepted XXX. Received YYY; in original form ZZZ}

\pubyear{2023}

\begin{document}
\label{firstpage}
\pagerange{\pageref{firstpage}--\pageref{lastpage}}
\maketitle

\begin{abstract}
We report on a $\rm{\ct}_{158\mu\rm{m}}$ search using the Atacama Large Millimeter/submillimeter Array (ALMA) on three lensed, confirmed {\lya} emitting galaxies at $z \sim 7$. Our targets are ultra-violet (UV) faint systems with stellar masses on the order of $M_{*} \sim 10^{9} M_{\sun}$.  We detect a single {\ct} line emission ($4\sigma$) from the brightest ($L \sim 2.4 \times 10^{10}L_{\odot}$) galaxy in our sample, MACS0454-1251. We determine a systemic redshift ($z_{\rm{\ct}} = 6.3151 \pm 0.0005$) for MACS0454-1251 and measure a {\lya} velocity offset of $\Delta v \approx 300 \pm 70 \rm{km\,s}^{-1}$. The remaining two galaxies we detect no {\ct} but provide $3 \sigma$ upper limits on their {\ct} line luminosities which we use to investigate the $L_{\textrm{\ct}} - \rm{SFR}$ relation. Overall our single {\ct} detection shows agreement with the relation for local dwarf galaxies. Our {\ct} deficient galaxies could potentially be exhibiting low metallicities ($Z<Z_{\odot}$). Another possible explanation for weaker {\ct} emission could be strong feedback from star formation disrupting molecular clouds. We do not detect continuum emission in any of the sources, placing upper limits on their dust masses. Assuming a single dust temperature of $T_{d}=35 \rm{K}$ dust masses ($M_{\rm{dust}}$) range from $< 4.8 \times 10^{7} M_{\odot} $ to $2.3 \times 10^{8} M_{\odot}$. Collectively, our results suggest faint reionization era sources could be metal poor and/or could have strong feedback suppressing {\ct} emission.



\end{abstract}




\section{Introduction}
In the past decade, the Atacama Large Millimeter/submillimeter Array (ALMA) observations of metal fine structure lines such as the $\rm{\ct}_{158 \mu \rm{m}}$ line have opened up studies into the epoch of reionization (EoR; $z>6$) by providing an unobscured view of galaxies. With the advent of the James Webb Space Telescope (JWST), which can also provide a similar a view of non-resonant optical lines (e.g., H$\alpha$), ALMA observations still provide a complementary view for far infrared (FIR) emission lines. FIR lines hold immense value for EoR studies because they are not affected by dust extinction, in contrast to rest-optical lines accessible with JWST. There have been numerous {\ct} detections in UV-bright, high-$z$ ($z>6$) galaxies \citep[e.g.,][]{Willott15, Carniani17, Laporte17, smit_rotation_2018, Matthee2019ApJ, Harikane19, Bakx20}. However, there are considerably fewer recorded {\ct} detections for characteristically faint, $L<L^{*}_{z \sim 7}$ (where $L^{*}$ is the characteristic luminosity) EoR galaxies \citep[e.g.,][]{Schaerer2015A&A, Watson15, Knudsen2016MNRAS, bradac_alma_2017, fujimoto_alma_2021, laporte2021MNRAS}. Faint galaxies are much more numerous than bright ($L<L^{*}_{z \sim 7}$) galaxies \citep[e.g.][]{Bouwens2022ApJ, Bolan2022MNRAS} and, as such, can be  a key driver of reionization. However, this connection has been heavily debated \citep[e.g.][]{Finkelstein2019ApJ, Naidu2020ApJ, Robertson2022, Endsley2023MNRAS}. In order to resolve this question, we need to study the physical properties of these fainter primordial systems; this step is key to understanding their role in cosmic reionization. 

Here we use {\ct} observations to study $z\sim 7$ galaxies. The {\ct} line is of particular interest because it  predominantly traces the dense neutral gas in photodissociation regions (PDRs, \citealp{Wolfire2022}) associated with molecular clouds, and the diffuse neutral gas \citep{Wolfire2003ApJ, Hollenbach1999}. It is the most luminous line in the FIR band ($\sim 0.1-1\%$ of the FIR luminosity, \citealp{BaierSoto2022A&A}) and one the strongest emission lines of star-forming galaxies at FIR/radio wavelengths \citep{carilli_cool_2013, stacey_158_1991, stacey_158_2010}. Additionally, the {\ct} line can be used to trace the systemic redshift, and therefore velocity, of the host galaxy because it is optically thin and not affected by dust extinction. When paired with detected {\lya} emission, we can estimate interstellar medium (ISM) properties of the galaxy under some assumptions. For instance, we can estimate the amount of neutral gas assuming the offset stems from {\lya} resonant scattering in ISM gas \citep{Mason2018ApJ}.\footnote{We remind that outflows might also generate comparable velocity offsets.} {\lya} photons travelling through neutral ISM scatter more frequently and emerge with a larger $\Delta v$ compared to photons traveling through less neutral ISM \citep{Yang2016ApJ,Yang2017ApJ,Guaita2017}. Previous observations of $z>5$ galaxies have typically measured $\Delta v \lesssim 500 \rm{km s}^{-1}$ \citep{Cassata2020A&A, carniani_2018MNRAS_kpc_gas_clump, pentericci_tracing_2016, Bunker2023, Prieto-Lyon2023arXiv_JWST} with the largest $\Delta v \approx 1000 \rm{km s}^{-1}$ recorded by \citet{BaierSoto2022A&A}. \citet{Cassata2020A&A} find that galaxies ($4.4<z<6$) with smaller $\Delta v$ have larger {\lya} rest-frame equivalent widths and $f_{\rm{esc}}(\textrm{\lya})$. However, for intrinsically faint systems ($M_{\rm{UV}} \gtrsim -20.5$) at $z>6$, a limited sample size restricts our ability to make robust conclusions (see \citealp{Endsley2022MNRAS} and references therein).

In addition to velocity offsets, we can also measure {\ct} line luminosities ($L_{\rm{\ct}}$) and evaluate the empirical {\ct} line luminosity to star formation rate ($\rm{L}_{\ct}-\rm{SFR}$) relation. This tight relation is  well established for a wide range of local galaxy types \citep[e.g.,][]{Boselli2002A&A, DeLooze2011MNRAS, DeLooze14, Sargsyan2012ApJ, Pineda2014A&A, Herrera2015APJ}, but initial {\ct} searches in $z>5$ galaxies revealed lower than expected or missing {\ct} emission from "normal" ($\rm{SFR}<100 \rm{M}_{\odot} \rm{yr}^{-1}$) star-forming galaxies (see \citealp{Carniani2020MNRAS} and references therein). The coined "{\ct} deficit" problem brought into question whether or not {\ct} could remain a good tracer of SFR at higher-$z$ systems as well as the applicability of the low-$z$ $\rm{L}_{\ct}-\rm{SFR}$ relation. Observational studies \citep[e.g.,][]{Maiolino15, Capak15, Knudsen2016MNRAS, matthee_alma_2017, carniani2018ApJ_himiko, Matthee2019ApJ, Schaerer_2020, Romano2022A&A} have focused on understanding the $\rm{L}_{\ct}-\rm{SFR}$ relation and if there indeed is missing {\ct} emission for $z\sim 7$. 

Theoretical studies have also extensively modeled the under/undetected {\ct} emission \citep[e.g.,][]{Vallini2013MNRAS, vallini2015ApJ,Olsen2017ApJ, Katz2017MNRAS, Katz2019MNRAS, Pallottini2017MNRAS,Pallottini2022MNRAS, Lupi2020MNRAS}. The analytical approach by \citet{Ferrara19} concluded that under-luminous {\ct} systems can result from a combination of factors: (a) large upward deviations from the Kennicutt-Schmidt relation (corresponding to bursty phases); (b) low metallicity; (c) low gas density, at least for the most extreme sources. These results have been supported by cosmological numerical simulations \citep{vallini2015ApJ, Pallottini19}. 

As mentioned previously, majority of EoR studies investigating {\ct} have targeted UV-bright galaxies, which traditionally have higher SFRs.   These larger and brighter galaxies are inherently easier to observe but do not exemplify galaxies at this epoch. Lensing, in our case by foreground galaxy clusters, serves as an excellent way to increase the apparent brightness and size of these fainter systems. In particular, the flux of such objects is increased by a magnification factor $\mu$ and resolution improved by $\sqrt{\mu}$. In this paper we report on a {\ct} study carried out with ALMA that targeted a set of lensed, sub-$L_*$ galaxies with SFRs<$20 M_{\odot} \rm{yr}^{-1}$. The lensed sizes of our galaxies are compact and smaller than the beam sizes of observations. All objects in our sample are spectroscopically confirmed EoR galaxies via detected {\lya} emission. We acknowledge a sample of galaxies all with detected {\lya} emission could potentially bias our results. However, recent results by Bolan et al. in prep find no significant difference in stellar UV properties (e.g, stellar mass, star formation rate, specific star formation rate,  UV magnitude, $\beta$-slope, age) between {\lya} emitters and non-emitters at $5<z<8.2$ in the similar stellar mass range to the one studied here.

   The paper is organized as follows. Section 2 explains various observations used to compile our sample and the specific ALMA data reduction performed on our data. In Section 3 we discuss the measurements and derived properties from the ALMA data as well as the spectral energy density (SED) fitting performed for galaxy property estimates. In Section 4 we analyze and discuss our results with literature findings. In Section 5 we summarize our conclusions. Throughout this paper we assume a $\Lambda$CDM concordance cosmology with $\Omega_m =0.27$, $\Omega_{\Lambda}=0.73$ and Hubble constant $H_{0} = 73 \rm{km\,s}^{-1} \; \rm{Mpc}^{-1}$. Coordinates are given for the epoch J2000.0, magnitudes are in the AB system, and we use the \citet{Chabrier2003PASP} initial mass function (IMF). 

\section{Observations and Data Reduction}
\begin{table*}
    \centering 
    \setlength{\tabcolsep}{2.5pt}
    \renewcommand{\arraystretch}{1.5}
    \caption{Properties of observed galaxies.}
    \begin{tabular}{|c|c|c|c|c|c|c|c|c|c|c|c|}
        \hline
        Target ID & $z_{\rm{Ly}\alpha}$ & $\mu_{\rm{best}}^a$ & Ref$^{b}$&   $\beta_{\rm{UV}}$    &   ${EW}_{\rm{Ly}\alpha}^b$    &   $M_{*}$    $\times f_{\mu}^{d,e}$ & $\rm{SFR}_{\rm{SED}}$  $\times f_{\mu}^{d,e}$ &  $\rm{SFR}_{\rm{UV}}$  $\times f_{\mu}^{d,f}$ & Age$^{e}$ &   $M_{\rm{UV}}^{d}$    & L$^{g}$\\ 
         -- & -- & -- &  -- &   --   &   \AA    & $10^{9} M_{\odot}$ & $M_{\odot} yr^{-1}$ & $M_{\odot} yr^{-1}$ & Myr & mag & $L^{*}$\\
        \hline    
        MACS0454-1251 & $6.323 \pm 0.002^{h}$ & $4.4 \pm 0.4$  & $[1]$ &  $-1.6 \pm 0.2^{b}$   &  $6.8 \pm0.9^{c}$   & $5.52_{-1.5}^{+1.6}$ & $13.5_{-3.9}^{+4.1}$& $8.82 \pm 0.15$ & $55_{-29}^{+58}$ &$ -20.8   \pm 0.1$ & $ 0.74    \pm 0.09$   \\
        MACS2129-1412 & $6.846 \pm 0.001$ & $11_{-0.7}^{+0.1}$ & $[2]$&  $-0.93 \pm 0.6^{e}$  &  $60 \pm 11$   & $1.73_{-0.61}^{+4.2}$ & $6.93_{-2.4}^{+5.9}$& $1.06 \pm 0.04 $ & $216_{-80}^{+330}$ & $-18.6 \pm 0.2 $ & $ 0.10   \pm 0.02$ \\
        RXJ1347-018 & $7.161 \pm 0.001$ & $21.4_{-1.3}^{+1.7}$ & $[3]$ &  $-1.6 \pm 1.0^{e}$   &  $27.2 \pm 6.5$   & $2.49_{-2.0}^{+3.0}$ & $ 7.5  _{-5.7}^{ +12  }$ & $0.29 \pm 0.01$ & $249_{-147}^{+335}$ & $-17.2 \pm +0.2$ & $0.03 \pm 0.01$ \\
    \end{tabular} 
    \\
    \flushleft{$^{a}$ To use a different magnification factor $\mu$, simply use $f_{\mu} \equiv \mu/\mu_{\rm{best}}$, where $\mu_{\rm{best}}$ is the magnification factor we adopt for each object.}
   \flushleft{$^b$ References for photometry, lens modeling and, \lya : $[1]$\citet{Huang2016ApJ_SURFSUP},$[2]$ \citet{Huang2016ApJ_2129}, $[3]$\citet{Hoag2019ApJ}.} 
   \flushleft{  $^{c}$ Value has been averaged from two night observations: $8.2\pm1.4 \textrm{\AA}$ (first night) and $5.4 \pm 1.0 \textrm{\AA}$ (second night).   }
   \flushleft{$^{d}$ Values have been corrected for lensing using $\mu_{\rm{best}}$.}
   \flushleft{$^{e}$ Values derived from SED fitting described in Section $3$.}
   \flushleft{$^{f}$ $\rm{SFR}_{\rm{UV}}$ calculated from lens-corrected $M_{UV}$ magnitudes with Eq.1 from \citet{kennicutt_star_1998} converted to Chabrier IMF via $0.63 \times \rm{SFR(Salepeter)}_{UV} =  \rm{SFR(Chabrier)}_{UV}$.}
   \flushleft{$^g$ We adopt the characteristic magnitude $M^{*} = -21.13 \pm 0.08$ from \citet{Bouwens2022ApJ}.}
   \flushleft{$^h$ Measured on two different DEIMOS slitmasks at $z = 6.32403$ and $z=6.32228$ with $\langle z \rangle = 6.3232 \pm 0.0018$. Uncertainty reflects difference between the mask measurements.}
\label{tab:Spec_table}
\end{table*}

\begin{figure}
        \includegraphics[width=0.5\textwidth]{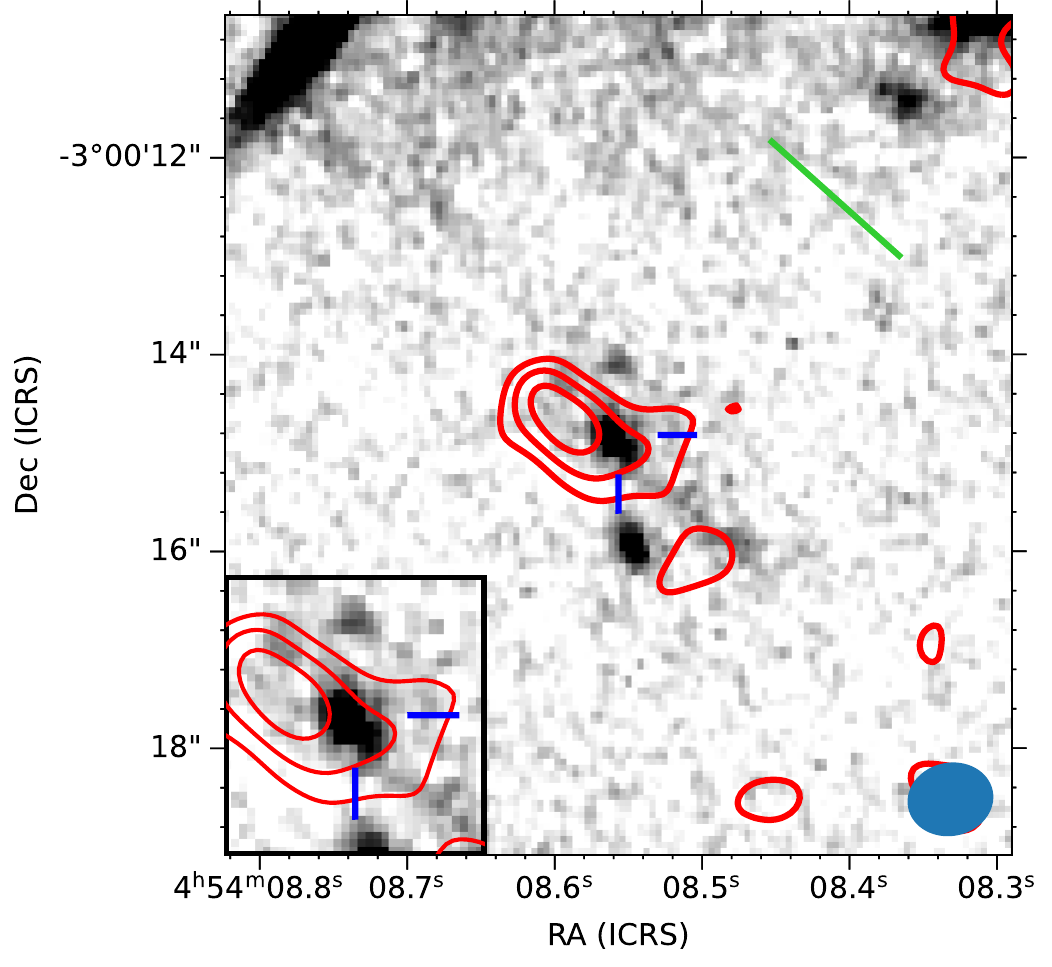}
    \caption{MACS0454-1251 velocity-integrated [CII] line intensity overlaid on a \emph{HST}/WFC3 F160W image. The contours are shown in red and are spaced linearly by intervals of $1\sigma$ which range from $2$, $3$, and $4\sigma$  ($\rm{RMS} = 87.1\rm{mJy/beam}$). The beam size ( $0\arcsec.88 \times 0\arcsec.82$) is given in the bottom right with a $2\arcsec \times 2\arcsec$  zoom-in shown in the bottom left.  The galaxy is located at an RA and DEC of $04$:$54$:$08.56$, $-03$:$00$:$14.82$ (UV centroid) with blue cross hairs marking the target in the center. The direction of shear is shown by the green line.  }
    \label{MACS0454_contour}
\end{figure}
The   lensing    galaxy clusters of our sample have been extensively studied in past works. All three clusters have imaging from the \emph{Hubble Space Telescope (HST)}. The Cluster Lensing and Supernova Survey with Hubble (CLASH, \citealp{Postman2012_CLASH}) program observed MACS2129 and RXJ1347. The two CLASH clusters were also spectroscopically observed with The Grism Lens-Amplified Survey from Space (GLASS ,   \citealp{Treu2015ApJ}) program. MACS0454 was imaged with the \emph{HST}-GO-11591(PI: Kneib)/GO-9836(PI:Ellis)/GO-9722(PI: Ebeling) programs. Additionally, the three clusters were imaged with the \emph{Spitzer} UltRa Faint SUrvey Program (SURFSUP \citealp{Bradac2014ApJ_surfsup}) which observed a total of ten lesning clusters. Spectroscopic follow ups using the DEep Imaging Multi-Object Spectrograph (DEIMOS, \citealp{Faber2003SPIE_DEIMOS}) and Multi-Object Spectrometer for InfraRed Exploration (MOSFIRE, \citealp{McLean2010_MOSFIRE}) instruments on \emph{Keck} confirmed {\lya} emission in the galaxies that make up our sample for this work \citep{Huang2016ApJ_SURFSUP, Huang2016ApJ_2129, Hoag2019ApJ}. 

ALMA observations (Proposal ID: 2019.1.00003.S) were carried out between March of 2020 and April 2021 using ALMA Band $6$ with 43 12-m antennae in array configurations C$43-4$ and C$43-5$. The precipitable water vapor (PWV) ranged from $0.5$mm to $2.1$mm. The spectral setup consisted of one spectral window centered on the expected observed frequency of $\rm{\ct}_{158 \mu \rm{m}}$ estimated from the {\lya} redshift. The remaining two spectral windows were used for continuum measurements. The on-source time varied from $20$ to $35$ minutes per target. 

The data was reduced and calibrated with the Common Astronomy Software Applications (CASA) package, version 6.1.1.15 following standard procedures. We reimaged the data with the CASA task   TCLEAN, adopting Briggs weighting with ROBUST=$2.0$ (effectively natural weighting) and adding UVTAPER=$0\arcsec.6$. We adopt the taper $= 0\arcsec.6$, given the typical {\ct} effective radius of $\sim 2 \rm{kpc}$ at $z\sim 4-6$ \citep{Fujimoto2020ApJ} which corresponds to $0\arcsec.3$.   The angular resolution for each object is equal to  their respective beam sizes which increased globally by a factor of $\sim 4$ when the taper was applied. The RMS also increased when the taper was applied by a factor of $\sim 1.56$ (MACS0454-1251), $\sim 1.36$ (MACS2129-1412), and $\sim 1.13$ (RXJ1347-018) respectively. The final beam sizes and RMS values are quoted in the captions of the contour figures.    All three targets were unresolved in their final data products and were spectrally binned into $25$-km s$^{-1}$ velocity channels. To ensure positional accuracy, we astrometrically calibrated \emph{HST} reference images to GAIA DR2 when comparing {\ct} to UV emission. Continuum maps were made with all spectral windows. 
\begin{figure}
	\includegraphics[width=\columnwidth]{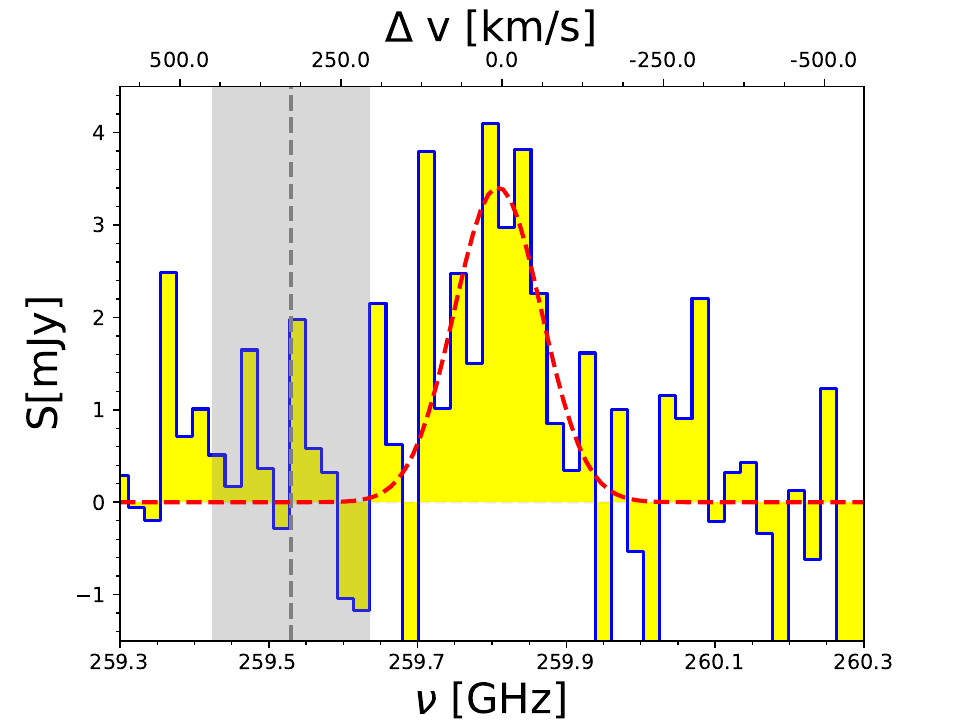}
    \caption{Extracted spectrum showing the detected {\ct} emission. Flux (left axis) shown as a function of frequency (bottom axis) and velocity (top axis). Red dashed line represents the best fit Gaussian and grey dashed line represents Lyman-$\alpha$ redshift with boxed $1\sigma$ uncertainty.}
    \label{fig:Glazer_spectrum}
\end{figure}

\begin{table*}
    \centering
    \setlength{\tabcolsep}{2.5pt}
    \renewcommand{\arraystretch}{1.5}
    \caption{Results from ALMA.}
    \begin{tabular}{|c|c|c|c|c|c|c|c|c|c|c|}
        \hline
        Target ID & $z_{\rm{\ct}}$ & $S_{\rm{line}}$ & $\rm{FWHM}$ & $S_{\rm{line}} \Delta v$ & $L_{\rm{\ct}} \times f_{\mu}$ & Continuum$^a$ & $L_{\rm{FIR}} \times f_{\mu}^{a,b}$ & $\rm{SFR}_{\rm{FIR}}^{a,b,c}$ & $M_{\rm{dust}}^{a,c}$\\ 
         -- & -- & $ (\rm{mJy})$ & (km$\rm{s}^{-1}$) & ($\rm{Jykm\,s}^{-1}$) & $(\times 10^{8} L_{\odot})$ & $(\mu \rm{Jy})$ & $(\times 10^{10} L_{\odot})$ & $M_{\odot} \rm{yr}^{-1}$ & $\times 10^{8} M_{\odot}$\\
        \hline    
        MACS0454-1251 & $6.3151 \pm 0.0005$ & $3.7_{-0.9}^{+0.9}$  & $163 \pm 46$ & $0.64 \pm +0.15$ & $1.5_{-0.4}^{+0.5}$ & $<1380$ & $<32$ & $<35$ & $<2.3$\\
        MACS2129-1412& $-$ & $-$ & $-$ & $-$ & $<0.043$ & $<18.3$ & $<0.19$ & $<0.21$ & $<0.93$\\
        RXJ1347-018& $-$ & $-$ & $-$ & $-$ & $<0.062$ & $<47.0$ & $<1.1$ & $<1.2$ & $<0.48$\\
    \end{tabular}
    \flushleft{$^{a}$ All limits are $3 \sigma$ }
    \flushleft{$^{b}$ Calculated from \citet{Kennicutt98_golabal_schmidt} Eq.3, which was the converted to Chabrier IMF via $0.63 \times \rm{SFR(Salepeter)}_{FIR} =  \rm{SFR(Chabrier)}_{FIR}$}
    \flushleft{$^{c}$ Assumed $T_d = 35 \rm{K}$.}
    
\label{tab:ALMA_results}
\end{table*}
\section{Measurements and Derived Properties}
We observed three highly magnified ($\mu \sim 4-20$) Lyman alpha emitting (LAEs) galaxies with ALMA at $z_{\rm{Ly}\alpha}=6.323 \pm0.003$ (MACS0454-1251), $6.846 \pm 0.001$ (MACS2129-1412), and $7.161 \pm 0.001$ (RXJ1347-018). We report a single {\ct} detection (4$\sigma$, Fig~\ref{MACS0454_contour}) from MACS0454-1251. We extracted the spectrum over an extended elliptical aperture (semi-axes: $\sim 1.\arcsec 5 \times 0.\arcsec 5$, pitch angle: $\sim 35^\circ$) to account of the slightly oblong shape in the {\ct} contours of the moment0 map (see Fig~\ref{MACS0454_contour}). We note that the SNR was slightly improved when the uvtaper was optimized to the elliptical morphology of the source. We fit a Gaussian (see Fig~\ref{fig:Glazer_spectrum}) to estimate the peak line flux $S_{\textrm{line}}=3.7\pm 0.9 \textrm{mJy}$. We calculate the integrated line flux ($S_{\rm{line,g}} \Delta v$) as $0.64 \pm 0.15 \rm{Jykm\,s}^{-1}$ with a FWHM of $163 \pm 46 \rm{km\,s}^{-1}$. We determine the systemic redshift $z_{\textrm{[CII]}}=6.315 \pm 0.001$ which is in close agreement with the redshift found via {\lya}  ($z_{\rm{spec}} = 6.323$). The calculated {\ct} line luminosity (eq.18, \citealp{Casey2014}) is $L_{\rm{\ct}}= 1.5^{+0.5}_{-0.4}\times 10^{8} L_{\odot}$ and find a velocity offset (the difference between {\lya} and {\ct}: $v_{\rm{Ly}\alpha}-v_{\rm{[CII]}}$) $\Delta v = 320 \pm 70\, \textrm{km\,s}^{-1}$.   We estimate the physical size of the emission by fitting a 2D Gaussian and deconvolving the clean beam from the map. The beam-deconvolved size of the major and minor axes are $1.37\arcsec \pm 0.18\arcsec$ and $0.41\arcsec \pm 0.21 \arcsec$. This corresponds to a physical size of $\sim 3.6 \pm 0.5 \rm{kpc} \times 1.1 \pm 0.6 \rm{kpc}$ after dividing by $\sqrt{\mu}$ to account for lensing. We divide the FWHM by $2.0$ to compare with effective radii found in a previous study \citep{Fujimoto2020ApJ} at $z=4-6$. The resulting {\ct} size of our semi-major axis ($\sim 1.8 \pm 0.3 \rm{kpc}$) is consistent with the $\sim 2-3 \rm{kpc}$ sizes reported in \citet{Fujimoto2020ApJ}.  

\begin{figure*}
\begin{minipage}{1.0\textwidth}
\noindent \begin{minipage}{0.5\textwidth}
\vspace{1cm}
\includegraphics[width=\textwidth]{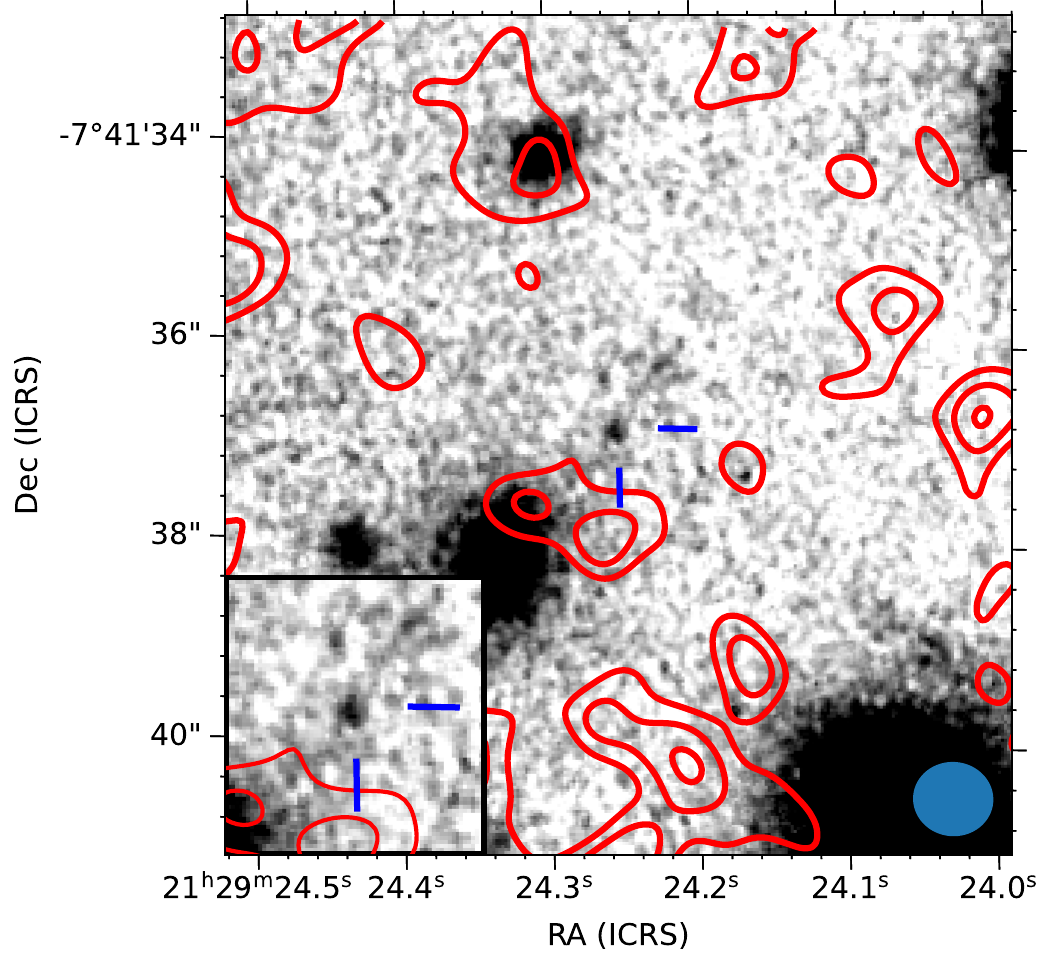}
\end{minipage}
\noindent \begin{minipage}{0.5\textwidth} 
\vspace{1cm}
\includegraphics[width=\textwidth]{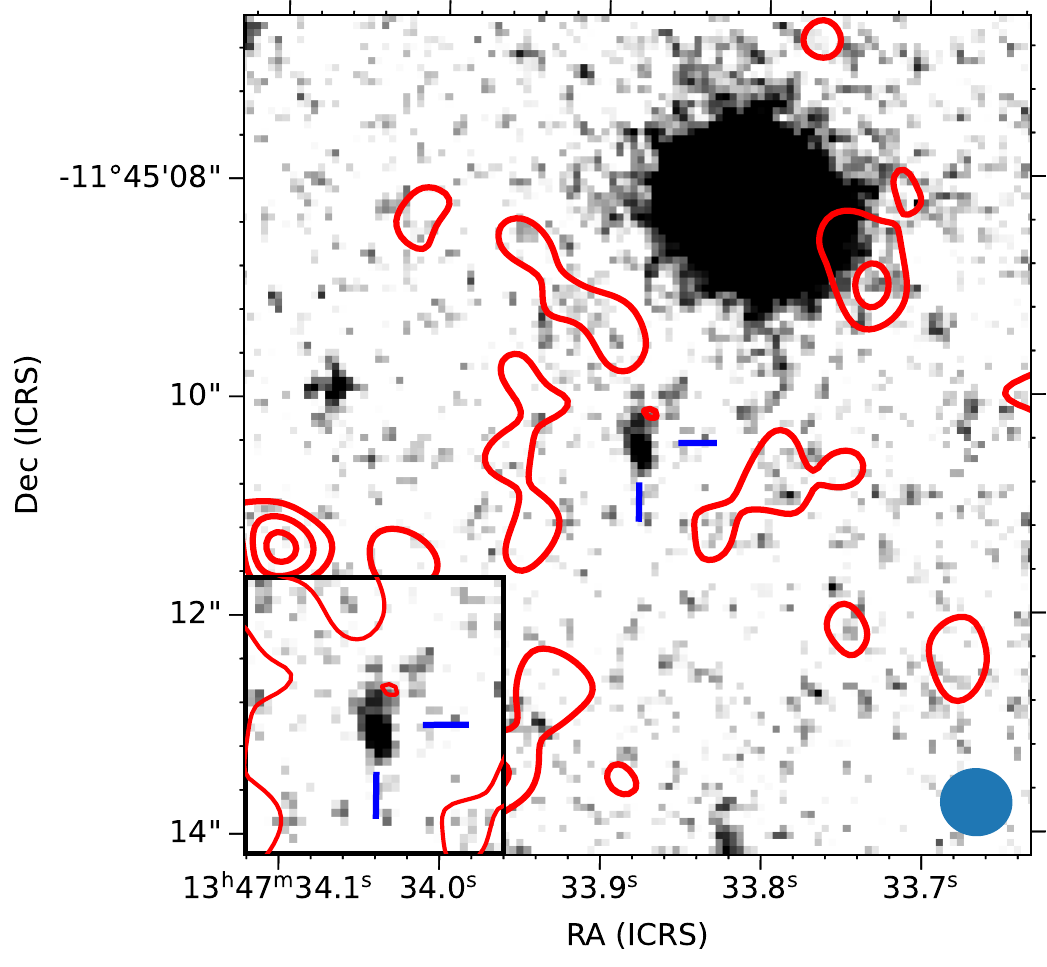}
\end{minipage}
\end{minipage}
\caption{MACS2129-1412 (left) and RXJ1347-018 (right) velocity-integrated [CII] line intensity overlaid on \emph{HST}/WFC3 F160W image. The contours are shown in red and linearly spaced at $1 \sigma$ intervals from $1-4\sigma$   ($\rm{RMS} = 15.3\rm{mJy/beam}$ and $34.8\rm{mJy/beam}$ respectively)   . The beam is given in the bottom right  ( $0\arcsec.82 \times 0\arcsec.76$ for MACS2129-1412, $0\arcsec.66 \times 0\arcsec.62$ for RXJ1347-018)    with a   $2\arcsec \times 2\arcsec$    zoom-in shown in the bottom left.   Respective RAs and DECs (for UV emission) of these two galaxies are $21$:$29$:$24.23$, $-07$:$414$:$35.96$ (MACS2129-1412) and $13$:$47$:$33.90$, $-11$:$45$:$09.40$ (RXJ1347-018) with the objects outlined by blue cross hairs on the \emph{HST} images.  }
\label{fig:contour_figure}
\end{figure*}

The remaining two galaxies, MACS2129-1412 and RXJ1347-018, yielded non-detections for {\ct} emission (see Fig~\ref{fig:contour_figure}). Because their intrinsic {\ct} line widths are unknown, we estimate their $L_{\rm{\ct}}$ ($3\sigma$) upper limits by integrating over the same width of channels ( $9$ channels, $225 \rm{km\,s}^{-1}$) as was done in MACS0454-1251. In the absence of known systemic redshifts, we use {\lya} redshifts to predict expected {\ct} emitting frequency of each object \footnote{This assumes there is no velocity offset between {\lya} and {\ct} emission for these two galaxies}. The RMS uncertainty was calculated using the $1\sigma$ intensity value extracted from a $0.\arcsec 5$ aperture centered to the HST imaging centroid of the target. We list the calculated $L_{\rm{\ct}}(3 \sigma)$ upper limits in Table \ref{tab:ALMA_results}.  

We constructed a continuum map for MACS0454-1251 from all four spectral windows (SPWs),  masking out the channels expected to contain {\ct} emission. As seen in Fig~\ref{fig:Glazer_spectrum}, the {\ct} line of MACS0454-1254 appears to be extended over roughly nine channels which corresponds to a width of $225 \rm{km\,s}^{-1}$. All nine channels were masked when producing the final continuum map from which we derived an upper limit on $L_{\rm{FIR}}$. We assume a wavelength range of $8 - 1000 \mu$m for $L_{\rm{FIR}}$ calculation. For the remaining non-detection galaxies, we constructed continuum maps in the same manner excluding nine channels centered on the central frequency. To estimate $L_{\rm{FIR}}$, we used a grey-body spectral energy distribution model \citep{Casey2012MNRAS}, assuming a spectral index of $\beta = 1.5$. In the absence of multi-band observations, we assumed a   uniform    dust temperature ($T_{\rm{d}} = 35 \rm{K}$) across all three objects and estimated $3 \sigma L_{\rm{FIR}}$ which is recorded in table \ref{tab:ALMA_results}.   In similar studies where dust continuum was not detected \citep{Fujimoto2022_z8_5_outflows, Fudamoto2023arXiv, Witstok2022MNRAS_dustcontinuum}, assumed dust temperatures varied $T_{\rm{d}}= 40 - 60 \rm{K}$. Other studies on mostly massive galaxies \citep{fujimoto_alma_2021, Bakx2021MNRAS_dusttemperaturez7, Algera2024MNRAS_colddust, Witstok2022MNRAS_dustcontinuum, Sommovigo2022MNRAS} also have a sizeable range on $T_{\rm{d}}= 32 - 59 \rm{K}$ with considerable error at low and high ends (note that employing single band techniques to estimate dust temperature was shown to overestimate temperatures by $\sim 15 \rm{K}$). Because $L_{\rm{FIR}}$ and $M_{\rm{dust}}$ are directly related to the assumed $T_{d}$, assuming a lower $T_{\rm{d}}$ sets a lower limit on $L_{\rm{FIR}}$ and a conservative upper limit on $M_{\rm{dust}}$.   

 Because we do not detect continuum in all three galaxies, we provide $\rm{SFR}_{FIR}$ and $M_{\rm{dust}}$ limits in Table ~\ref{tab:ALMA_results}.    We estimated $\rm{SFR}_{\rm{FIR}}$ using eq.3 from \cite{Kennicutt98_golabal_schmidt} and converted from Salpeter IMF to Chabrier IMF. We assumed a dust mass absorption coefficient of $\kappa = \kappa_{0}(\nu/\nu_{0})^{\beta_{\rm{d}}}$, where $\kappa_{0} = 0.232 \rm{m}^2 \rm{kg}^{-1}$ at $\nu_{0} = 250\mu \rm{m}$ \citep{Draine2003, Bianchi2013}.  

To estimate age and stellar mass ($M_{\rm stellar}$) of the sample, we use Bayesian Analysis of Galaxies for Physical Inference and Parameter EStimation (BAGPIPES, \citealp{Carnall2018MNRAS}). BAGPIPES fits physical parameters using the MultiNest sampling algorithm \citep{Feroz2008MNRAS, Feroz2009MNRAS}. We use the default set of stellar population templates from Bruzal and Charlot \citep[BC03]{Bruzual2003MNRAS}. The SED fitting is done assuming the \citet{Kroupa2001_imf} IMF which we convert to \citet{Chabrier2003PASP}, a metallicity of 0.02$Z_{\odot}$, the Calzetti dust law \citep{Calzetti2000ApJ}, and a constant star formation history. We allow dust extinction to range from $A_{v} = 0-3$ magnitudes. The $M_{\rm stellar}$ values reflected in Table~\ref{tab:Spec_table} have been converted to Chabrier IMF via conversion factor $0.923$. We take the general prescription for the fitting from \citet{Strait2020ApJ_properties}. See Bolan et al. in prep. for more information on the SED fitting. 

\section{Results}
\subsection{Velocity and Spatial Offset}
\label{sec:Velocity Offset}
Observational studies have used {\ct} emission as a tracer of systemic velocities \citep[e.g][]{pentericci_tracing_2016, Matthee2019ApJ}. Using the peak {\ct} emission from the extracted spectrum of MACS0454-1251, we find $\Delta v_{\rm{Ly}\alpha-\rm{[CII]}} \approx 320 \pm 70 \rm{km\,s}^{-1}$. This is shown in Fig~\ref{fig:Glazer_spectrum} where the velocity axis is centered on the {\ct} emission and the grey dashed line represents Ly$\alpha$. The magnitude of the offset falls within literature ranges quoted for $z=2\sim3$ \citep{Erb2014ApJ} and high-$z$ galaxies  \citep[e.g.,][]{Cassata2020A&A, Endsley2022MNRAS}. 

The spectrum shows a clear redshift of {\lya} emission. Given the resonant nature of {\lya}, revealing the direct cause of the offset is both a complex and active place of research. Redshifted {\lya} could indicate scattered emission from outflowing (or expanding) gas from the galaxy. Outflows may originate from strong star formation feedback which could reduce the covering fraction of neutral gas in the ISM and boost {\lya} escape \citep{Jones2013ApJ, Trainor2015ApJ, Leethochawalit2016ApJ}. {\lya} could also be redshifted by neutral hydrogen inside a galaxy's ISM, where the emerging $\Delta v$ would be a proxy for the column density of neutral hydrogen \citep{Yang2016ApJ, Yang2017ApJ, Guaita2017}. Additionally, the model put forth by \citet{Mason2018ApJ} used $\Delta v$ as a way to measure the intergalactic medium (IGM) neutral fraction. A more neutral IGM will scatter {\lya} photons more causing Ly$\alpha$ to emerge at a higher $\Delta v$ relative to systemic. 

It is not uncommon for $z>5$ galaxies to have {\ct} emission tracing the systemic redshift of the galaxy but spatially offset from the UV component \citep{Maiolino15, Willott15, Capak15, Carniani17, carniani2018ApJ_himiko, carniani_2018MNRAS_kpc_gas_clump, Jones_2017, Matthee2019ApJ, Fujimoto2022_z8_5_outflows}. In fact, spatial offsets between {\lya} and UV have also been found at $z>5$ \citep{Hoag2019MNRAS.488..706H, Lemaux2021MNRAS.504.3662L}. \citet{carniani_2018MNRAS_kpc_gas_clump} shows that most of the {\ct} spatial offsets are indeed physically motivated but further observations are needed to understand the mechanisms causing the offsets. Recently, \citet{Fujimoto2022_z8_5_outflows} was able to determine the necessity of past outflow activity in a galaxy at $z \sim 8.5$ based on dual observations with ALMA and JWST. 

We roughly estimate the {\ct}-UV spatial offset in MACS0454-1251 using the brightest pixels found in the {\ct} moment0 map and the centroid of the \emph{HST} rest-UV image. The lensed spatial offset is $\sim 0.\arcsec 5$  which is greater than the ALMA astrometric accuracy of $\sim 0.\arcsec 08$ \footnote{Calculated from ALMA technical handbook \url{https://almascience.nrao.edu/documents-and-tools/cycle10/alma-technical-handbook}.}. Taking into account the magnification, we estimate the lens-corrected offset to be $\sim 1.4 \mbox{kpc}$ \footnote{For isotropic lensing distortion, we spatially scale with $ / \sqrt{\mu}$}. The offset could be physically associated with intrinsic ISM properties (e.g. different distribution in the ionized vs neutral gas phase). A possible explanation for the spatial offset could also be the ejection of material by galactic outflows or galaxy mergers \citep[e.g.,][]{Maiolino15, vallini2015ApJ, Pallottini2017MNRAS,Katz2017MNRAS,Gallerani2018MNRAS, Kohandel2019MNRAS}.

\subsection{$L_{\rm{\ct}}-\rm{SFR}$ Relation}
\begin{figure}
	\includegraphics[width=\columnwidth]{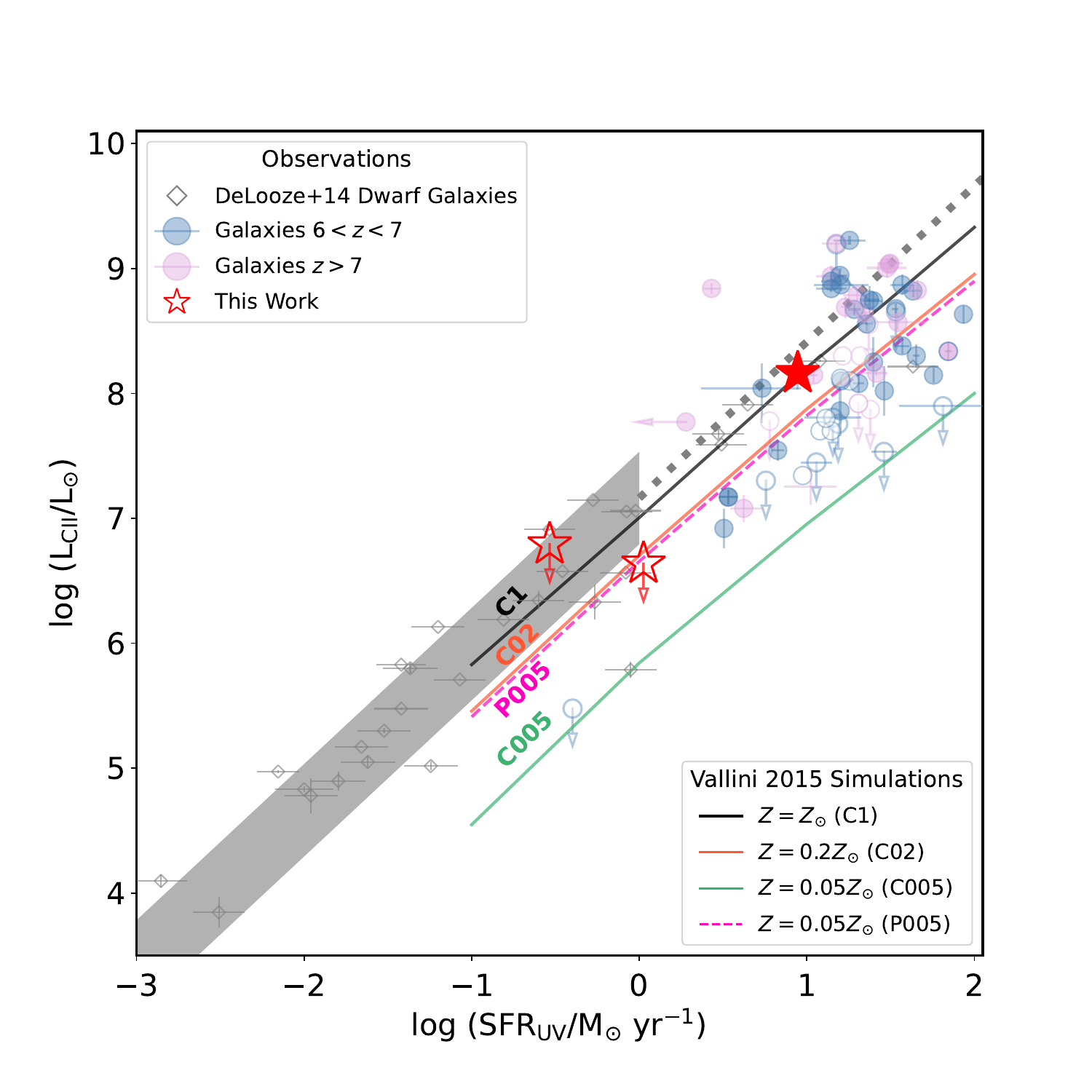}
    \caption{A plot of $L_{\textrm{\ct}}$ vs. $\rm{SFR_{\rm{UV}}}$ relation for our study (red stars) and literature. The blue ($z>6$) and purple ($z>7$) circles show previous studies \citet{Willott15, Maiolino15, Knudsen2016MNRAS, pentericci_tracing_2016, matthee_alma_2017,bradac_alma_2017,carniani_2018MNRAS_kpc_gas_clump, Smit18,Hashimoto2018Natur,Matthee2019ApJ, Schouws22a, harikane_large_2020, fujimoto_alma_2021, Molyneux2022MNRAS, Schaerer2015A&A, Inoue16, Carniani2020MNRAS, Watson15,Wong2022, Katz2017MNRAS, Bakx20, Bowler2018MNRAS, Hashimoto2019PASJ, Fujimoto2022_z8_5_outflows, Heintz2023ApJ, Ferrara2022MNRAS, Sommovigo2022MNRAS, Valentino2022ApJ}. Upper limits are denoted by open markers. The lines represent the results from \citet{vallini2015ApJ} with black $Z=Z_{\odot}$ (C1), orange $Z=0.2Z_{\odot}$ (CO2), and   green    $Z= 0.05Z_{\odot}$ (C005) representing a constant metallicity. The magenta dashed line (P005) represents a density-dependent metallicity of $Z=0.05Z_{\odot}$.   The best-fit dwarf relation from \citet{DeLooze14} is shown in gray with measurements as open gray diamonds. Above $\rm{SFR} =1 M_{\odot} / yr$, the relation is shown as a gray dotted line with $1\sigma$ scatter to emphasize that only a few galaxies constrain the higher $L_{\rm{\ct}}/\rm{SFR}$ regime.  } 
    \label{fig:Glazer_SFR-LCII}
\end{figure}

\label{sec:LCII-SFR}
In Fig~\ref{fig:Glazer_SFR-LCII} we show the $L_{\rm{\ct}}-\rm{SFR}_{\rm{UV}}$ relation for our galaxies (stars) alongside available $z>6$ observations from the literature. The reported $\rm{SFR}_{\rm{UV}}$ values for our objects can be found in Table \ref{tab:Spec_table} along with $\rm{SFR}_{\rm{SED}}$ values that were derived through SED fitting described in Section 3. $\rm{SFR_{\rm{UV}}}$ is calculated assuming no dust attenuation and therefore should be considered a lower limit as it does not include obscured star formation. We also recognize that in the event of a more top heavy IMF (i.e., the median of the mass-to-light ratio of stars being born decreases), $\rm{SFR}_{\rm{UV}}$ would be an overestimate assuming constant dust properties. 

Our single {\ct} detection, MACS0454-1251, falls within the $1\sigma$ scatter of the \citet{DeLooze14} relation for dwarf ($\sim 0.2$ dex below). We note, however, MACS0454-1251 is the most UV luminous galaxy ($L=0.74L^{*}$) in our sample. The detection is also consistent with the C1 model put forth by the \cite{vallini2015ApJ} simulations which corresponds to a galaxy with constant solar metallicity. If we compare this to our non-detections, the upper limit set by RXJ1347-018 also is consistent with the \citet{DeLooze14} relation; having a scatter of $\sim 0.3$ dex from the average. The upper limit set by MACS2129-1412 is not consistent with the \citet{DeLooze14} relation, falling $\sim 0.6$ dex from the dwarf galaxies. The one displayed \cite{vallini2015ApJ} model it could possibly support is the C005 model which represents a galaxy with a constant metallicity of $Z = 0.05Z_{\odot}$. 

 {The analysis presented here focuses on the $\rm{SFR_{UV}}$ component, but $\rm{SFR_{SED}}$ as well as upper limits on $\rm{SFR_{FIR}}$ are provided for completeness. The $\rm{SFR_{UV}} / \rm{SFR_{FIR}}$ limits are not very constraining for our galaxies given we only have upper limits on $\rm{SFR_{FIR}}$ making us unable to exclude any model considered in this work. We do note the $\rm{SFR_{FIR}}$ limit for MACS2129-1412 shows $\rm{SFR_{FIR}} < \rm{SFR_{UV}}$. For the $M_{\rm{dust}}/M_{*}$, all three of our galaxies showed limits $\log M_{\rm{dust}}/ \log M_{*} \la 0.8$. Our limits are consistent within $1\sigma$ of the overall values reported in \citet{Sommovigo2022MNRAS} ($\log M_{\rm{dust}}/ \log M_{*} \approx 0.74 - 0.81$) who studied $13$ higher-mass galaxies at $z\sim7$. Our limits are also consistent with the dust build up reported in \citet{Pozzi2021A&A_dustmass}}.  

  We also advise caution when drawing comparisons between our objects and the dwarf relation in \citet{DeLooze14}. As shown in Fig ~\ref{fig:Glazer_SFR-LCII}, majority of dwarf galaxies (gray diamonds) used to make the best-fit relation (gray dotted line) are much lower luminosity systems than those reported in this paper as well as literature. Above $\rm{SFR}\sim 1 \rm{M_{\odot}yr^{-1}}$ the dwarf relation is less constrained, being driven by only a handful of dwarf galaxies.

  The solar and sub-solar metallicity models by \citet{vallini2015ApJ} provide a unique view on the metal content of our galaxies. The possibility of a sub-solar metallicities is not surprising for our three low-mass galaxies and aligns with recent work \citep{Curti2023arXiv, Nakajima2023ApJ}. Although \cite{Curti2023arXiv} probed a moderately lower mass range than our sample, they found their $z>6$ galaxies exhibiting sub-solar metallicities ($Z < Z_{\odot}$), with a scaling relation drawn near $\sim0.1Z_{\odot}$. A more comparable study by \cite{Nakajima2023ApJ} reported  $Z < Z_{\odot}$ for a sub-sample of galaxies ranging from $z=6-10$ with a somewhat larger range of masses.    While we do not have the observations to determine the true metallicities of our sample, a low metallicity could explain the absence of {\ct} in our non-detections and would not contradict their respective upper limits. Another possible reason for the lack of detected {\ct} emission could be negative feedback disrupting molecular clouds (MCs, \citealp{vallini2015ApJ}). For example, {\ct} emission predominately originates from PDRs. Negative feedback disrupting MCs would reduce the PDR layer at the edge of the MC where most of the {\ct} emission comes from. Additionally, we know stellar feedback is efficient in galaxy centers, places where there is very active star formation. In fact, studies done at lower-$z$ have supported a connection between feedback efficiency and SFR surface density \citep[e.g.,][]{Hayward10.1093/mnras/stw2888, Heckman2011ApJ}. Since high-$z$ galaxies are more compact than local analogs, this would also support negative feedback suppressing {\ct} emission at the systemic redshift of the galaxy.

\section{Conclusions}
We present new ALMA observations investigating {\ct} $158 \mu\rm{m}$ line emission in three lensed galaxies at $z>6$. We find a $4\sigma$ {\ct} detection in our most luminous galaxy ($L \sim 0.7L^{*}$) MACS0454-1251 and calculate the systemic redshift to be $z_{\rm{\ct}}=6.3151 \pm 0.0005$. We measure a $\Delta v = 320 \pm 20 \rm{km\,s}^{-1}$ and calculate $L_{\textrm{\ct}} = 1.5_{-0.4}^{+0.5} \times 10^{8}L_{\odot}$. For the remaining two galaxies we do not detect {\ct} emission but provide $3 \sigma$ upper limits for their $L_{\textrm{\ct}}$. Our main findings are:

(i) MACS0454-1251 exhibits both a velocity and a spatial offset between between ALMA {\ct} and rest-frame HST UV emission. The spatial offset is larger than the the ALMA astrometric uncertainty which could mean the offset is physically motivated by intrinsic ISM properties. 

(ii) Feedback is a very important process in both the emission of {\ct} and {\lya}. On one hand, strong feedback can suppress {\ct} emission by disrupting MCs \citep{vallini2015ApJ}. On the other hand strong feedback from star formation can drive outflows which redshifts the {\lya} line. A possibility that puts together the spatial offset and the {\ct} deficit is that feedback destroys the emitting MCs at the center, allowing only displaced MCs to contribute to the {\ct} emission.

(iii) Our single {\ct} detection in MACS0454-1251 falls within the $1\sigma$ scatter of the \citet{DeLooze14} $L_{\rm{\ct}}-\rm{SFR}$ relations. As such, it would support with the applicability of the $L_{\rm{\ct}}-\rm{SFR}$ relation put forth by \citet{DeLooze14} for $z > 6$ galaxies. That being said, the upper limits set by our RXJ1347-018 and MACS2129-1412, would argue an inconclusive result, especially with the $3 \sigma$ upper limit of MACS2129-1412 still falling well below \citet{DeLooze14}. In general, more observations of low-mass, UV faint galaxies are needed in order to break this degeneracy. 

(iv) Low metallicity is a possible justification for fainter galaxies falling below the \citet{DeLooze14} $L_{\textrm{\ct}}-\textrm{SFR}$ relation \citep{Vallini2013MNRAS, Ferrara19}. Based on recent work \citep{Curti2023arXiv}, we would expect lower mass, ($M_{\rm{stellar}} \lessapprox 10^{9.5}$) galaxies at $z>6$  to exhibit sub-solar metallicites. A possible scenario for {\ct} deficient systems with very low metalicity  is powerful feedback; involving the destruction of star forming sites occurring during bursty evolutionary phases in relatively chemically unevolved systems  \citep{vallini2015ApJ, Ferrara19}. 

\section*{Acknowledgements}


 We acknowledge support from the program HST-GO-16667, provided through a grant from the STScI under NASA contract NAS5-26555. MB also acknowledges support from the ERC Advanced Grant FIRSTLIGHT and Slovenian national research agency ARRS through grants N1-0238 and P1-0188.  RLS acknowledges support provided by NASA through the NASA Hubble Fellowship grant \#HST-HF2-51469.001-A awarded by the Space Telescope Science Institute, which is operated by the Association of Universities for Research in Astronomy, Incorporated, under NASA contract NAS5-26555. AF acknowledges support from the ERC Advanced Grant INTERSTELLAR H2020/740120.

\section*{Data Availability}
The original data used in the work can be found and downloaded from the ALMA archive \url{https://almascience.nrao.edu} using the science project ID: 2019.1.00003.S. The reduced data generated in this research will be shared on reasonable request to the
corresponding author.



\bibliographystyle{mnras}
\bibliography{Carbon_Emission_at_z7.bib}








\bsp	
\label{lastpage}
\end{document}